# Automated Bioacoustic Monitoring for South African Bird Species on Unlabeled Data

Michael Doell[1],* Dominik Kuehn[1],* Vanessa Suessle[1,2,3]*
Matthew J. Burnett[2] Colleen T. Downs[2] Andreas Weinmann[3] Elke Hergenroether[1]

*joint first authors
[1]Department of Computer Science, University of Applied Sciences Darmstadt, Schoefferstrasse 3, Darmstadt, Germany
[2]Centre for Functional Biodiversity, School of Life Sciences, University of KwaZulu-Natal, P/Bag X01, Scottsville, Pietermaritzburg, 3209, South Africa
[3]Algorithms for Computer Vision, Imaging and Data Analysis Group, University of Applied Sciences Darmstadt, Schoefferstrasse 3, Darmstadt, Germany
@ michael.doell@extern.h-da.de, dominik.kuehn@extern.h-da.de, vanessa.suessle@h-da.de

## ABSTRACT
Analyses for biodiversity monitoring based on passive acoustic monitoring (PAM) recordings is time-consuming and challenged by the presence of background noise in recordings. Existing models for sound event detection (SED) worked only on certain avian species and the development of further models required labeled data. The developed framework automatically extracted labeled data from available platforms for selected avian species. The labeled data were embedded into recordings, including environmental sounds and noise, and were used to train convolutional recurrent neural network (CRNN) models. The models were evaluated on unprocessed real world data recorded in urban KwaZulu-Natal habitats. The *Adapted SED-CRNN* model reached a F1 score of 0.73, demonstrating its efficiency under noisy, real-world conditions. The proposed approach to automatically extract labeled data for chosen avian species enables an easy adaption of PAM to other species and habitats for future conservation projects.

## 1 INTRODUCTION

The monitoring of wildlife is essential for wildlife conservation and biodiversity management. While camera traps were the tool of choice for monitoring many species [1, 2], the practicality of relying solely on visual data diminished for certain species because of factors such as body size and behavior [3, 4]. Passive acoustic monitoring (PAM) emerged as an alternative to collect large datasets [5], especially for avian species. The recorded soundscapes gave vital information on the ecosystems, its population and biodiversity [6, 7, 8]. Sound meters enabled a constant and unobtrusive recording of ambient soundscapes, collecting large amounts of data [9]. The manual analyses of such data created a bottleneck for research and the presence of noise complicated the analyses [10, 11]. The classification of avian species using acoustic data is further complicated by factors such as inter-species similarity, overlapping calls and diverse characteristics of intra-species calls.

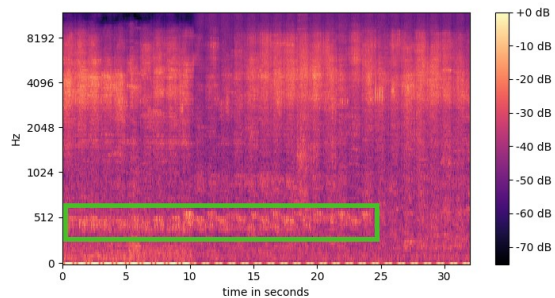

Figure 1: Example of a red-eyed dove Mel-spectrogram with a high level of background noise.

Avian species might have different prominence in the soundscape depending on the intensity and frequency of their calls [6, 8]. The automated analyses of data from camera traps, using artificial intelligence (AI) based methods, has been intensively researched over the past years [12].

Transfer of such methods for audio recordings is still in its infancy compared with camera traps [9]. The conversion of audio recordings into spectrograms facilitated the application of AI-based computer vision methods [13, 14]. With an "image" as input, detection and classification models could be applied to the converted audio recording. But the lack of comprehensive labeled datasets caused, that existing models often exhibit species specificity, limited geographical cover-age or required re-training and programming expertise to be applied to new case studies [14, 9, 4]. To the best of our knowledge, available ready-to-use models have not been developed to process large datasets or they required a certain amount of manual labeling.

In this work we developed a framework that can automatically detect the presence of certain avian species without the requirement of prior manual labeling for unprocessed data recorded in noisy environments. This framework exhibits a high degree of generality for a simple extension to other species in the future.

## 2 RELATED WORK

PAM has been applied to a variety of species, including marine fauna, birds, insects, terrestrial mammals and amphibians [15, 16], leading to a growing research field based on audio data. The analyses of such data presented numerous challenges, depending upon the species, their habitats, and the various types of sound pollution associated with them. Insights derived from this research are valuable for informing economic decision-making processes. For instance, wildlife occurrence assessments can be crucial in areas surrounding wind turbines to ensure the safety of endangered species like birds and bats [17, 18]. Numerous studies have focused on bats due to their reliance on sound for orientation and the informative nature of their calls [16]. Bat calls were detected by converting audio recordings to Mel-frequency cepstral coefficients (MFCCs) and processing them with convolutional neural networks (CNNs) [19]. The identification of bird calls has increasingly attracted research interest as well, prompting the initiation of public challenges like *BirdClef* [20, 21], which provided essential data to accelerate the expansion in avian bioacoustics research. The datasets were focused on species of the northern hemisphere. Existing deep learning models like *Nighthawk*, specialized for short-duration calls, were retrained on the Merlin Sound ID framework with datasets manually labeled for in-flight vocalizations of pre-selected nocturnal migratory species [22]. While existing applications like *BirdNET* [23] offered user-friendly interfaces for bird call identification, they were not designed to handle large datasets and frequently relied on location-specific species recognition. Further research is necessary for species in other geographic locations. Especially for rare species, that are often of special interest, the data availability is sparse [15].

One approach in the field of audio analyses involved treating audio data as a computer vision problem. Audio signals were converted into Mel-spectrograms, which then could be processed using CNNs [24]. The CNN models, which were pretrained on ImageNet, were fine-tuned using labeled audio files, which significantly improved bird call classification and demonstrated their ability to handle the complexity of bioacoustic data [25, 26]. However, generating sufficient labeled data for such analyses is labor and challenging [27]. Data augmentation is an option to efficiently generate labeled datasets. Approaches to achieve labeled images for anomaly detection in an industrial context [28] resonated with our approach to obtain training data by overlaying labeled bird sounds onto background noise, circumventing the scarcity of labeled data.

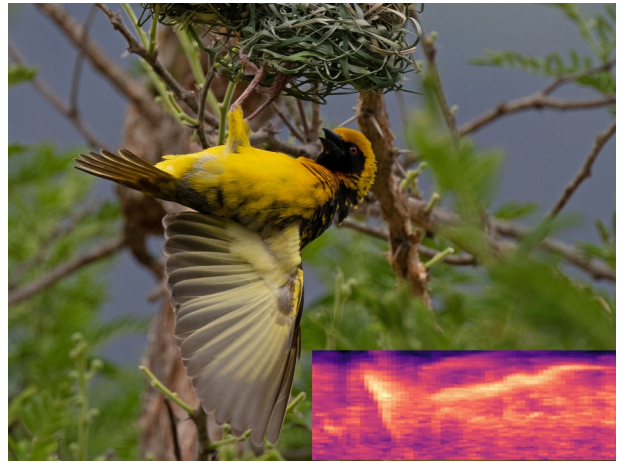

Figure 2: Village weaver while nest building and a spectrogram of its call.

The study [29] that compared models that capture temporal dependencies with those that do not, made findings that hybrid models, incorporating explicit temporal layers, significantly outperformed ImageNet-based models in the classification of bird acoustic data. This comparison underscored that the presence of temporal layers, including mechanisms like long short-term memory (LSTM), gated recurrent units (GRU) in recurrent neural networks (RNN) led to more accurate outcomes.

## 3 DATASET AND ITS GENERATION

We generated a dataset by automatically combining unlabeled and labeled data from existing sources to train our models. The fundamental dataset consisted of unlabeled audio recordings in which we embedded snippets of labeled data records.

### 3.1 Data Sources

We utilized 27 unlabeled audio recordings originating from a PAM program at different locations in suburban areas in the KwaZulu-Natal (KZN) Province, South Africa. The dataset was recorded with *Wildlife Acoustics - Song Meter Mini 2 AA*[1] at all locations. Each recorded file was saved as a *.wav* file with a length of about one hour. On the contrary, the labeled training data were exported from the *Xeno-Canto*[2] database. This is a publicly accessible collection of bird vocalizations from various global regions contributed by ornithologically interested volunteers. Other than the KZN dataset, the samples were recorded by different devices. Each recording on *Xeno-Canto* is labeled with the heard species and a rating on the sound quality from *A* (indicating the highest quality) to *E* (denoting the lowest quality).

---

[1] https://www.wildlifeacoustics.com/products/song-meter-mini-2-aa

To enhance the robustness and generalization capability of our model, we included labeled data across all quality ratings for the training dataset. Recordings for six chosen species were selected (Table 1).

| Class | Latin name | No. of Files |
|---|---|---|
| Brown-hooded Kingfisher | *Halcyon albiventris* | 70 |
| Dark-capped Bulbul | *Pycnonotus tricolor* | 346 |
| Hadada Ibis | *Bostrychia hagedash* | 177 |
| Olive Thrush | *Turdus olivaceus* | 49 |
| Red-eyed Dove | *Streptopelia semitorquata* | 129 |
| Village Weaver | *Ploceus cucullatus* | 134 |

Table 1: Labeled data records exported from *Xeno-Canto* per selected bird species.

### 3.2 Combining Labeled and Unlabeled Data

For the generation of the training dataset, files from the unlabeled KZN dataset were used as background and the labeled samples extracted from the *Xeno-Canto* were added to the background file. More precisely, randomly selected labeled samples from the *Xeno-Canto* subset (Table 1) for each of the species of interest were inserted at random locations in the KZN background audio file creating overlaps (Figure 3).

The generated dataset was automatically labeled based on where the samples were embedded into the background file. Audio files can be labeled with different approaches. The labels mark the start and end point of a call based on time. We chose one-dimensional labels for this implementation, where each call was defined by its start and end point. Using the one-dimensional approach increased the range of possibilities to use labeled data from existing platforms, like *Macaulay*[3] and *Xeno-Canto*, for future applications and species. The start and end points of the embedded overlapping samples defined the label for that occurrence of the species. For each KZN background file, the labels were provided as a two-dimensional list, where every occurrence of a species was saved with the start and end point (Figure 3). This process was carried out for a total of 27 KZN files. An equal distribution of files per species were inserted, to cope with the problem of unbalanced datasets. The *fill density* parameter regulated the amount of *Xeno-Canto* audio files that were integrated into a KZN audio file and therewith the relation between labeled bird calls and ambient noise. An exception is the *max fill density* setting: here all available labeled data for the relevant species were embedded in the KZN background files, resulting in a high density of labeled *Xeno-Canto* embeddings, but imbalanced distribution of species. Inspired by the natural behaviour of multiple birds chirping simultaneously, multiple em-bedded samples of the same or different species could be overlapping (Figure 3). The generated dataset finally consisted of around 27 hours of the KZN background soundscape with a maximum of 905 embeddings from *Xeno-Canto*.

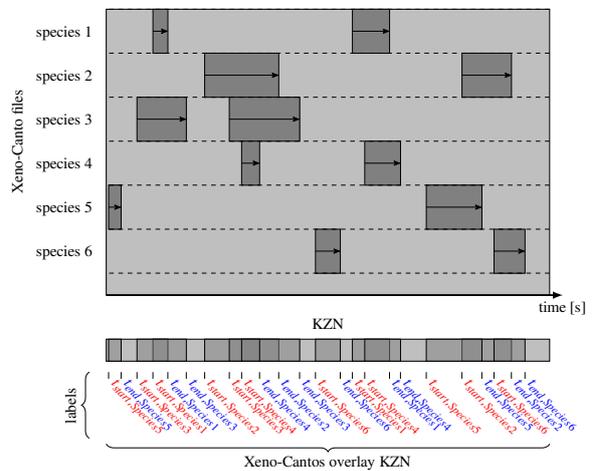

Figure 3: Passive acoustic monitoring data record (KZN) enriched with randomly sampled labeled data snippets: Selecting *k* items for each species and overlaying them at randomly selected timesteps in the KZN background record. Bottom: Darker shades of grey imply a higher overlapping.

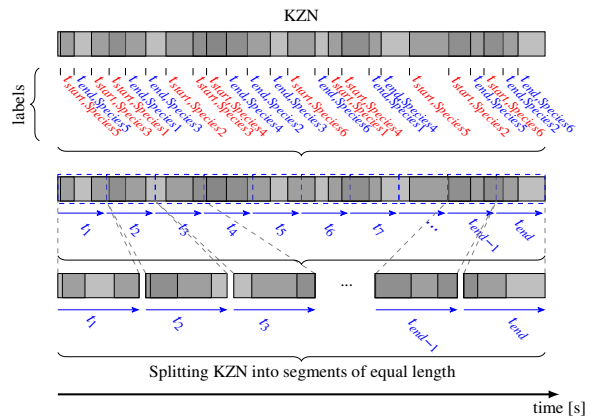

Figure 4: Example of splitting randomly sampled labeled data into segments of equal length of 1 second.

---
[2] https://xeno-canto.org
[3] https://www.macaulaylibrary.org

### 3.3 Label Formatting

For an unambiguous automatic labeling and an effective training of the neural network, it was crucial to maintain a uniform format. The recorded KZN dataset varied in length and therefore needed to be split into equally long labeled subsequent segments. In our implementation each segment had a length of one second, which would eventually divide a call into multiple segments of one second as shown in Figure 4. Each segment then had an individual binary label for the presence/absence of each species within that time interval $t_i \in [0, t_{end}]$. Start and end points of calls were rounded to the closest second and assigned to all segments that lay in that interval.

### 3.4 Transformation into Mel-spectrograms

The waveform files from the generated dataset were converted into spectrograms, providing the advantage of separating temporal and spectral information, allowing better feature learning [30].

Each segment was transformed from a 1D-audio waveform into a 2D-Mel-spectrogram [31]. Thereby the x-axis represented the time in seconds (s) and the y-axis the frequencies in Hertz (Hz) on the Mel-scale. The colour intensity depicted the magnitude in decibels (dB) (Figure 5). By stacking each Mel-spectrogram in depth, the output data would get reshaped from multiple 2D-Mel-spectrograms into a 3D-shaped cube (Figure 5). The fixed segment length and the predefined frequency range ensured the uniformed shape of the input image, fitting the model's requirements.

### 3.5 Impact of Noise

Our work aimed at providing a tool to predict the presence of certain bird species from audio files collected in natural habitats. The soundscape of such recordings encompassed a variety of sounds, including anthropophony (human-made sounds), biophony (sounds from non-human organisms), and geophony (natural sounds like wind and rain) [32].

We generated a labeled training dataset, which included different types and intensities of noise, to train our models. This approach offered the dual benefit of preserving the inherent noise within the original dataset and also adding labels for the model training, enabling a robust model training process.

Our unlabeled KZN dataset as well as our labeled *Xeno-Canto* subset included a wide range of different noises, including biophonetic sounds from a broad spectrum of non-target animal species, like insects, canines (barking dogs, *Canis lupus familiaris*) and various non-categorized avian species (geese, chickens), as well as environmental or geophonetic sounds like wind, rain, flowing streams and vegetative movements (rustling leaves). Additionally, anthropogenic noises such as human voices, mechanical noises (e.g. machinery or technical equipment) and motor sounds (cars, motorbikes and trains) were present in part of the audio files. The KZN dataset, recorded in suburban environments had a high noise pollution from anthropogenic origin, especially in the recordings of the urban neighbourhood area.

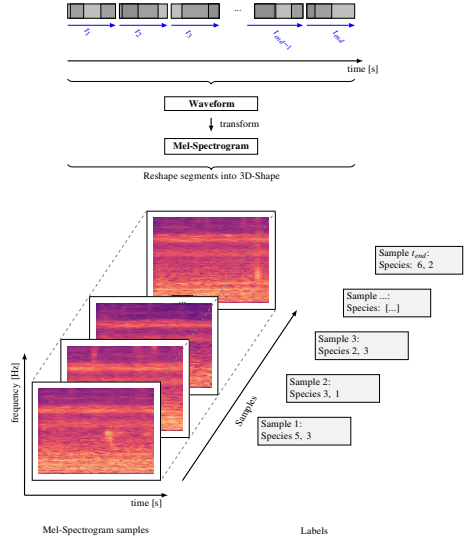

Figure 5: Sequence of reshaping the 2D-Mel-spectrograms into 3D-shaped format to ensure a fixed segment length.

The complexity and variability in noise types obscuring the bird calls of interest, presented a significant challenge for the processing and analysis of the audio data. For machine learning applications, noise could impede the model from identifying and learning relevant features [33]. For the robustness and the ability of the model to adapt for practical use in audio signal processing in diverse and noisy realworld environments, where noise was an inevitable factor, different types of noise had to be considered. By using the KZN samples as a background, the model was confronted with the noise present in the application case during the training process.

## 4 THE MODELS

The neural network was trained to learn features in the Mel-spectrograms to detect the chosen bird species calls from a soundscape and predicts the species' presence or absence. Because of the overlapping *Xeno-Canto* and KZN background audio data, the model was confronted with variable types of noise during training to simulate a natural environment. Unlabeled bird calls in the original KZN data could not be considered in the training process and were treated as noise.

### 4.1 Employed Architectures

We utilized convolutional recurrent neural networks (CRNNs), grounded in their widespread recognition as an effective approach for sound event detection (SED) [34]. In particular, we focused on a foundational CRNN model architecture, referred to as *SED-CRNN* [31]. While the *SED-CRNN* in the original study [31] utilized 40 Mel bands as input, we explored an adaptation of this model, named *Adapted SED-CRNN*, where we increased the number of Mel bands to 128, as suggested in their work. Additionally, we employed the *SELDnet* framework [35], omitting its localization component to align with our focus solely on sound event detection.

The initial stage of these CRNN architectures included convolutional blocks, each consisting of a sequence of a convolutional layer, batch normalization, rectified linear unit (ReLU) activation and a max pooling layer, for feature extraction from spectrogram images. Following the convolutional blocks, bidirectional gated recurrent unit (GRU) [36] layers were integrated, which analyzed data sequences in both forward and reverse directions, addressing temporal dynamics. The temporal outputs from the GRU layers were directly fed into dense layers for the final classification task [31, 35]. The model architecture is schematically shown in Figure 6. Sigmoid activation was used in the output layer, paired with binary cross-entropy as loss function [37]. Afterwards a threshold-based binarization step was applied to convert probabilistic outputs into distinct class predictions.

### 4.2 Training the Models

In our experiments, we trained and compared *SED-CRNN*, featuring 40 Mel bands against the *Adapted SED-CRNN* and *SELDnet*, both utilizing higher-dimensional 128 Mel bands. For *SED-CRNN* and the *Adapted SED-CRNN* models, we selected a temporal input window of 5 seconds. The *SELDnet* model required a longer 32 second input window because of additional architectural constraints. Each model processed single-channel inputs. The specifications of the input hyperparameters for each model are summarized in Table 2.

| Model | Mels | Segment Length |
|---|---|---|
| *SED-CRNN* | 40 | 5s |
| *Adapted SED-CRNN* | 128 | 5s |
| *SELDnet* (w/o loc.) | 128 | 32s |

Table 2: Input hyperparameters of the employed model architectures.

All models were trained with a batchsize of 4 and a default of 300 epochs on the generated files combining the KZN dataset and the *Xeno-Canto* subset. Overfitting was mitigated using *EarlyStopping* [38], which terminated training upon halting of loss improvements, resulting in variable epoch count of less than 300 (Table 3). We avoided transfer learning to isolate the impact of our embedding-based pre-processing method. This approach, coupled with non-filtered spectrograms, allowed a direct assessment of how this method performed in bird call classification.

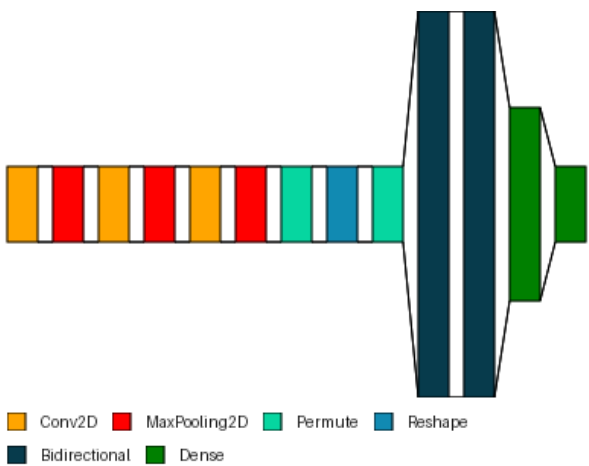

Figure 6: Example of a basic CRNN architecture with different types of layers. All models were based on the this basic structure, but with varying input segments length and number of Mel bands.

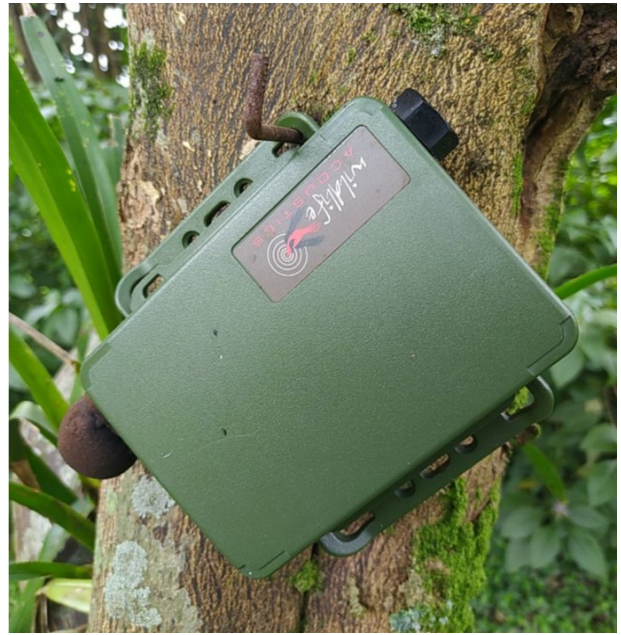

Figure 7: Audio recorder setup in habitat for passive acoustic monitoring.

| Model | Fill Density | Epochs | Precision | Recall | F1 Score | Loss | Accuracy |
|---|---|---|---|---|---|---|---|
| *SED-CRNN* | 10 | 42 | 0.34 | 0.46 | 0.40 | 0.30 | 0.91 |
| | 50 | 39 | 0.49 | 0.50 | 0.50 | 0.28 | 0.92 |
| | max | 39 | 0.55 | 0.45 | 0.49 | 0.21 | 0.95 |
| *Adapted SED-CRNN* | 10 | 35 | 0.56 | 0.74 | 0.63 | 0.17 | 0.95 |
| | 50 | 31 | 0.67 | 0.80 | 0.73 | 0.14 | 0.96 |
| | max | 31 | 0.64 | 0.77 | 0.70 | 0.14 | 0.96 |
| *SELDnet (w/o loc.)* | 10 | 48 | 0.32 | 0.67 | 0.43 | 0.35 | 0.90 |
| | 50 | 80 | 0.62 | 0.77 | 0.69 | 0.18 | 0.95 |
| | max | 43 | 0.35 | 0.76 | 0.48 | 0.38 | 0.89 |

Table 3: Performance comparison of models based on *fill density* at a threshold of 0.5. *Max fill density* defines that all available labeled data samples for the relevant species were embedded in the background file.

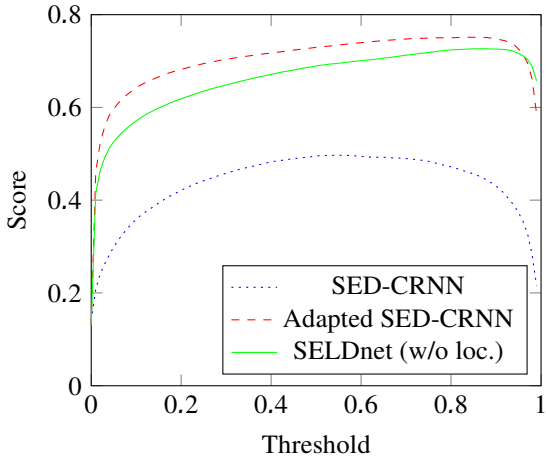

Figure 8: Comparison of F1 scores across different binarization thresholds trained with a fill density of 50.

## 5 RESULTS

In the evaluation of our models, we primarily focused on precision, recall, and the F1 score, because of the imbalanced nature of our dataset which featured a significantly high proportion of negative samples. We chose precision and recall for their focus on true positives and F1 score to balance these metrics to effectively assess the model performance amidst predominantly occurring negatives. Accuracy might yield misleading results in such contexts, as a model biased towards the majority class could still achieve a high score [39]. However, we have included accuracy in Table 3 to provide a baseline overview of overall model performance and model effectiveness in correctly classifying all samples, both positive and negative.

### 5.1 Performance of the Models

For a consistent evaluation of all models, we set a threshold of 0.5 to determine the binary classification results, as shown in Table 3. Across all models, we observed a consistent increase in precision, recall, and F1 score for a *fill density* between 10 to 50 samples per KZN background file. At *max fill density*, there was a small decrease in recall and F1 score, indicating an optimal performance around medium *fill density*. The *SED-CRNN* model, characterized by its low-dimensional frequency input, exhibited a unique precision pattern compared with the other models. *SED-CRNN* showed a steady increase in precision, rising from 0.34 at a *fill density* of 10 to 0.55 at *max fill density*, while the remaining models exhibited a slight decrease in precision at *max fill density*. A comparative analysis of F1 scores across the spectrum of binarization thresholds at a *fill density* of 50 (Figure 8), which was the most effective in the evaluation (Table 3). The *Adapted SED-CRNN* model reached a maximum F1 score of 0.73 at a threshold of 0.84. The *SELDnet* model attained its highest F1 score of 0.69 at a threshold of 0.87, indicating a slightly delayed peak performance relative to the *Adapted SED-CRNN*. The *SED-CRNN* model reached its maximum F1 score of 0.5 at a threshold of 0.56.

### 5.2 Performance in Real-World Application

An unseen validation dataset from the KZN dataset was used as a real world application to validate the framework. The data were recorded in a permanent long-term monitoring, recording continuously 24 hours a day at different locations, capturing different types and levels of noise. Two validation files from two different habitats, one in a suburban neighbourhood and one in a botanical garden with 40 minutes of duration in total. The occurrences of the species: Brown-hooded kingfisher (*Halcyon albiventris*), dark-capped bulbul (*Pycnonotus tricolor*), hadada ibis (*Bostrychia hagedash*), olive thrush (*Turdus olivaceus*), red-eyed dove (*Streptopelia semitorquata*) and village weaver (*Ploceus cucullatus*) were manually labeled by an expert. In the two habitats, different levels of noise were present in the recordings. Both included other species, traffic and biophonic noise, while the data recorded at the neighbourhood had more background noise in terms of human voice and machinery. The detection threshold was set to 0.1 to catch a maximum of potential bird call detections. The best detected bird species, measured by the F1 score, was the brown-hooded kingfisher in the suburban neighbourhood recording with a score of 0.80 (model: *SELDnet*) and a precision of up to 1 (model: *SED-CRNN*), meaning that all brown-hooded kingfisher detections predicted by that model were correct. No brown-hooded kingfisher was present in the botanical garden and therefore no direct comparison was feasible. The best performing model on the present species for the botanical garden reached a F1 score of 0.45 for the dark-capped bulbul (model: *SELDnet*) and a maximum precision for the red-eyed dove (model: *SED-CRNN*). The performance on the raw data was generally lower than on the generated dataset with the embedded *Xeno-Canto* data. The good performance for the dark-capped bulbul species could be reasoned by the highest amount of available training data for this species (Table 1).

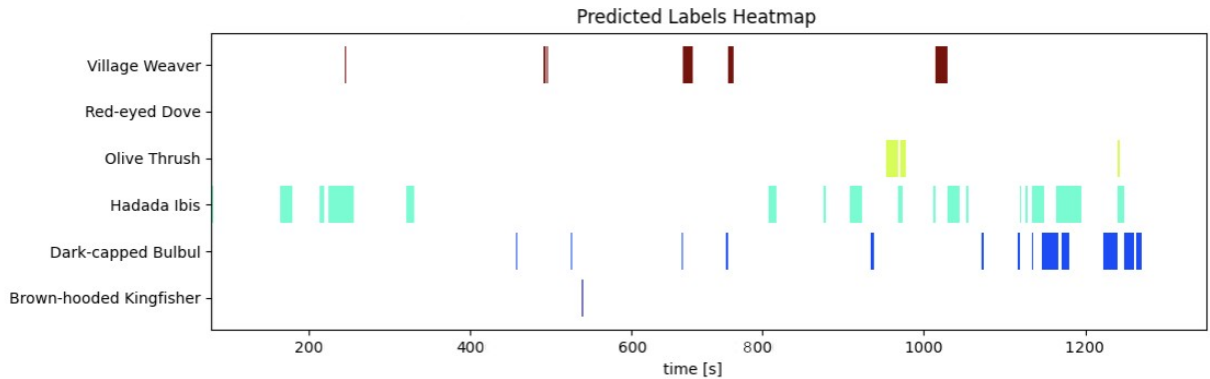

Figure 9: Sample GUI export for predictions. The presence of each species in each time interval is highlighted. Predictions of calls from different species may overlap.

The best performing model over all species, measured by the F1 scores, was the *SELDnet* with *max fill density* reaching an F1 score of 0.27 on the botanical dataset and 0.25 on the neighbourhood dataset. The lower value for the neighbourhood recording could be reasoned by the higher level of noise.

Typical misclassifications included children's voices incorrectly classified as hadada ibis. Misclassifications could be caused by the similarity of calls from species that have not been included in the training process or background noises in the *Xeno-Canto* data that were misleadingly learnt by the model as features for a species. The hardest species to classify and validate was the village weaver. At the location of the neighbourhood recording, village weaver nests were present, resulting in nearly constant village weaver callings, whereas at the botanical garden no nests were present and no village weaver called, leading to zeros in the evaluation metrics for that species.

The varying distribution of species, or even their absence at specific recording locations, as well as the proximity of nests to recording sites, posed challenges to the evaluation process.

## 6 CONCLUSIONS

We developed a robust framework for the detection of certain bird species from noisy PAM data without the requirement of manual labeling. The framework offers a pipeline to extend and train for additional species, not limited to avian species.

The selection of the species might have an impact on the performance of the models, because some species have more prominent calls and therefore were easier to detect and distinguish in spectrograms analyzed by humans and models. The setting of the detection threshold for the model prediction affected the detection rate. Depending on the application, there was a trade-off between prioritizing only secure detections and thereby overlooking less certain ones and wanting all potential detetctions, bearing the risk of having false detections.

In this study we faced the issue of unknown false negative examples. The background noise of the KZN dataset might include calls of the bird species of interest, but were not labeled. More labeled data, also for the background files might improve further model training.

For further research we plan to focus on adding more bird species or even species of other classes like amphibians, mammals or other animals producing sounds. The existing GUI, that was presently only used for testing purposes, could be further improved to be used as a tool by research groups (Figure 9). We plan to implement and train further model architectures besides the *SED-CRNN* and *SELDnet* into the framework. Furthermore, an approach with transformer models will be conducted to test, if timely dependencies in bird calls can be captured more effectively for better classifications [40].

## 7 ACKNOWLEDGMENTS

The project was made feasible thanks to the assistance of a scholarship provided by the DAAD (German Academic Exchange Service).